\documentclass{article}

\usepackage{microtype}
\usepackage{graphicx}
\usepackage{subfigure}
\usepackage{booktabs} 

\usepackage{hyperref}

\usepackage[accepted]{icml2023}

\usepackage{amsmath}
\usepackage{amssymb}
\usepackage{mathtools}
\usepackage{amsthm}

\usepackage[capitalize,noabbrev]{cleveref}

\theoremstyle{plain}

\theoremstyle{definition}

\theoremstyle{remark}

\usepackage[textsize=tiny]{todonotes}

\begin{document}

\twocolumn[
\icmltitle{Mobile Internet Quality Estimation using Self-Tuning Kernel Regression}



\icmlsetsymbol{equal}{*}

\begin{icmlauthorlist}
\icmlauthor{Hanyang Jiang}{yyy}
\icmlauthor{Henry Shaowu Yuchi}{yyy}
\icmlauthor{Elizabeth Belding}{xxx}
\icmlauthor{Ellen Zegura}{zzz}
\icmlauthor{Yao Xie}{yyy}
\end{icmlauthorlist}

\icmlaffiliation{yyy}{H. Milton Stewart School of Industrial and Systems Engineering, Georgia Institute of Technology, Atlanta, Georgia, USA}
\icmlaffiliation{xxx}{Department of Computer Science, University of California Santa Barbara, California, USA}
\icmlaffiliation{zzz}{School of Computer Science, Georgia Institute of Technology, Atlanta, Georgia, USA}

\icmlcorrespondingauthor{Yao Xie}{yao.xie@isye.gatech.edu}
\icmlkeywords{Machine Learning, ICML}

\vskip 0.3in
]



\printAffiliationsAndNotice{\icmlEqualContribution} 

\begin{abstract}

Modeling and estimation for spatial data are ubiquitous in real life, frequently appearing in weather forecasting, pollution detection, and agriculture. Spatial data analysis often involves processing datasets of enormous scale. In this work, we focus on large-scale internet-quality open datasets from Ookla. We look into estimating mobile (cellular) internet quality at the scale of a state in the United States. In particular, we aim to conduct estimation based on highly {\it imbalanced} data: Most of the samples are concentrated in limited areas, while very few are available in the rest, posing significant challenges to modeling efforts. We propose a new adaptive kernel regression approach that employs self-tuning kernels to alleviate the adverse effects of data imbalance in this problem. Through comparative experimentation on two distinct mobile network measurement datasets, we demonstrate that the proposed self-tuning kernel regression method produces more accurate predictions, with the potential to be applied in other applications.

\end{abstract}

\section{Introduction}
Since the invention of cellular-network-based wireless internet access more than two decades ago, the mobile internet has become an integral component of daily life, enabling a myriad of activities across diverse spheres, including communication, entertainment, commerce, and education. While urban areas generally boast robust mobile internet connectivity due to extensive infrastructure, rural areas often experience compromised connectivity \cite{vogels2021some}. This disparity is largely due to the lack of sufficient cell towers and infrastructural limitations, driven by geographical dispersion, lower population density, and reduced economic incentives for investment in these regions. Precisely finding these areas can lead to better planning of networks \cite{niu2021greening}.

Identifying these underserved or unserved areas with poor connectivity is vital for telecommunication companies and policymakers \cite{mangla2021tale}. By targeting these regions, service providers can strategically enhance infrastructure, improving coverage and service quality. On a broader scale, comprehensive connectivity mapping can inform policy-making, ensuring digital inclusivity and narrowing the digital divide between urban and rural areas. Improving mobile internet connectivity in rural areas also holds the potential for significant socio-economic development, contributing to regional growth and inclusivity. Thus, recognizing and addressing regions with poor mobile internet connectivity is a crucial step toward promoting digital equity and advancing societal progress. In the United States, the Broadband
Equity, Access, and Deployment (BEAD) program will allocate substantial resources for expansion in the next few years, guided in part by available data regarding spatial quality~\cite{BEAD}.

In this work, we examine the question of predicting mobile internet quality based on an in-depth analysis using one of the largest open datasets, namely the Ookla dataset \cite{ookla2022}, which captures mobile internet connectivity measurements worldwide. Users conduct speed tests on their internet connections using Ookla's tools. When a user initiates a speed test, data such as the user's IP address, the Internet Service Provider's (ISP) identity, and the connection's speed (both download and upload measured connection bandwidth) are logged. However, like many real-world datasets, it presents certain challenges. It is characterized by a relatively high variation of data and a conspicuous absence of measurements in specific areas.

To approach these issues, we develop a novel self-tuning bandwidth kernel regression method.
The hallmark of our method is its flexibility in the spatial imbalance of data. In contrast to conventional models, our approach is designed to adaptively determine the kernel regression bandwidth for a location. More specifically, we use larger kernel regression bandwidth where the data gets sparser. This adaptive capability enables our model to more effectively handle the varying data density and spatial distribution inherent in our dataset.

In addition, we are cognizant of the computational resources required to process large quantities of data. To mitigate this issue, we incorporate techniques specifically designed to reduce the computational cost of our model, making it more efficient for real-world deployment. The robustness and effectiveness of our proposed method are substantiated by a comprehensive comparison with two baseline models: Gaussian Process regression \cite{rasmussen2006gaussian} and the basic kernel regression \cite{parzen1962estimation}. This comparative analysis is carried out across two distinct datasets, offering further insight into the performance and adaptability of our method.

In the domain of data-driven mobile internet quality prediction, there are two principal approaches currently: geospatial interpolation \cite{molinari2015spatial,riihijarvi2018machine,tripkovic2021cluster,eller2021propagation,chakraborty2017specsense} and supervised machine learning \cite{alimpertis2019city,rozenblit2018machine,eller2022deep}. Geospatial interpolation employs sparse measurements, treated as anchoring points, to generate dense signal-strength maps in their immediate vicinity. Kriging is used in these works \cite{molinari2015spatial,riihijarvi2018machine,eller2021propagation}. \cite{molinari2015spatial} compares the performances of Kriging and points out that regular Kriging is fairly robust. \cite{eller2021propagation} develops a propagation-aware Gaussian process regression considering factors like the block of buildings. While these techniques are conducive to estimating performance in existing networks, they offer limited applicability in network planning as adjustments can only be studied post-deployment. Furthermore, barring a few exceptions, these methodologies do not inherently account for the effect of the propagation environment.

On the other hand, supervised machine learning methods have also been explored \cite{enami2018raik,wu2020artificial,rozenblit2018machine,alimpertis2019city}. Some techniques \cite{alimpertis2019city} primarily focus on creating local performance maps, utilizing absolute coordinates or cell identifiers. However, these approaches lack cross-area generalization capabilities. There exist alternative supervised strategies \cite{shoewu2018fuzzy,jo2020path} that restrict themselves to area-independent features, like the distance between the user equipment and base stations. Nevertheless, in numerous cases, any observed error reductions predominantly arise from fitting the distribution of a specific measurement campaign rather than enhancing the efficacy of traditional methods at a macroscopic level.

In this study, we aim to create an estimation of mobile internet quality that covers an entire state, not just urban areas as many previous studies have done. Some works including \cite{adarsh2021coverage} realize the importance of this field, but they focus more on the data collection and analysis part instead of developing methods. The lack of data from rural areas makes this a tough challenge, setting our work apart from existing approaches. We also show that our method performs much better than Kriging which is commonly used in geospatial interpolation work. The results of our work can guide active sampling in underserved areas, helping to improve our study further.

The rest of the paper is organized as follows. We first discuss the literature relevant to cellular network connectivity prediction. Then we give a detailed description of the datasets used in the paper. After that, we introduce the methods used in this paper. Finally, we show the numerical experiments on the two different datasets.

\section{Data Description}
The dataset utilized in our study is sourced from the open datasets provided by Ookla. The data gathered by Ookla from 2019 to 2022 encapsulates the performance metrics of mobile internet connections for a multitude of users worldwide. Key variables in this dataset include geographical coordinates (longitude and latitude), mean download speed (MB/s), mean upload speed (MB/s), count of tests conducted in each area (aggregated for user privacy into  $600m^2$ grid blocks), the number of distinct devices utilized for testing, and a comprehensive score assessing the connection speed, among other variables.

\begin{figure}[t]
    \centering
    \includegraphics[scale=0.315]{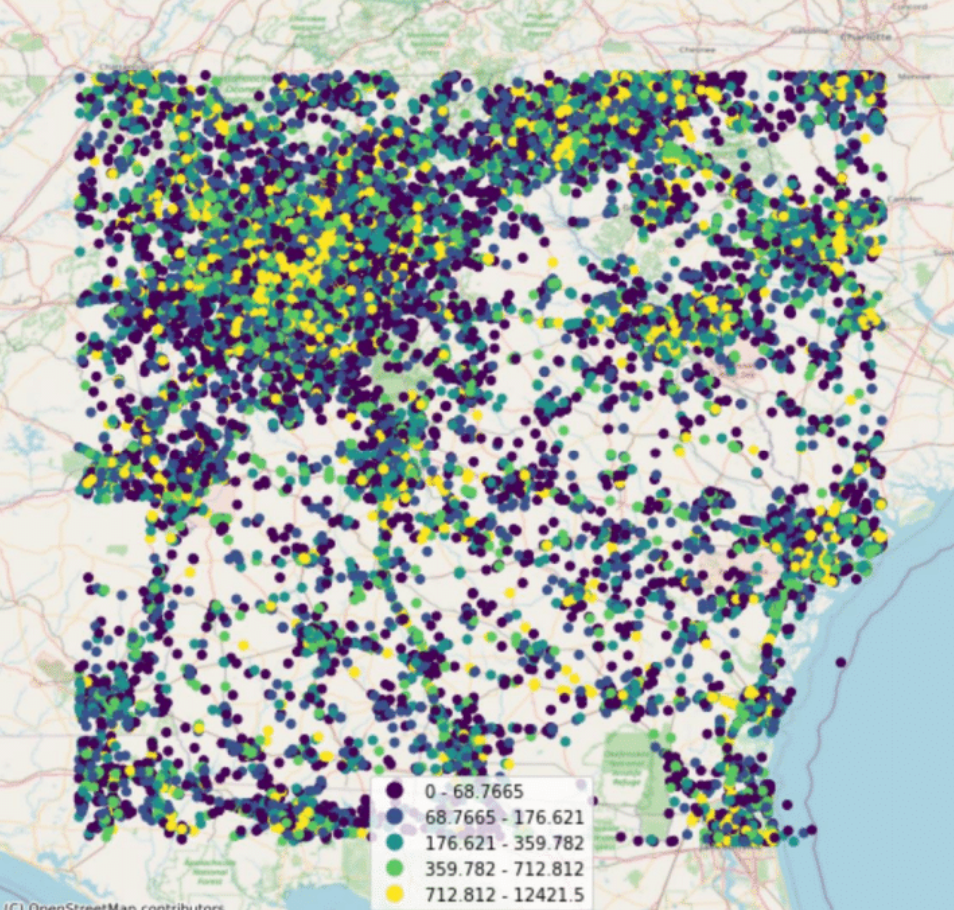}
    \includegraphics[scale=0.385]{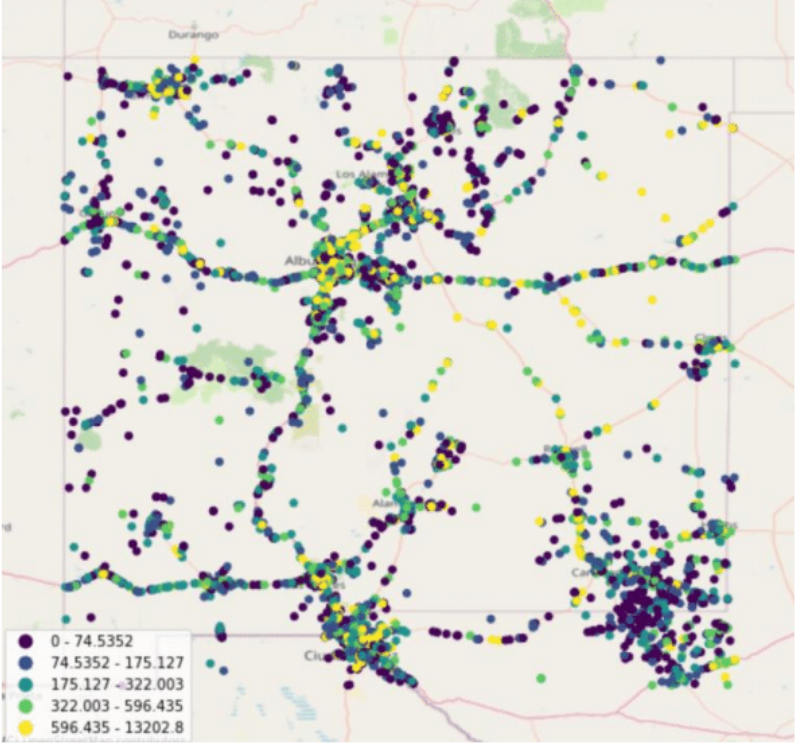}
    \caption{Mobile connection score of Georgia and New Mexico, where the data in Georgia is very dense, and in New Mexico is sparse.}
    \label{s1}
\end{figure}

\begin{figure}[t]
    \centering
    \includegraphics[scale=0.315]{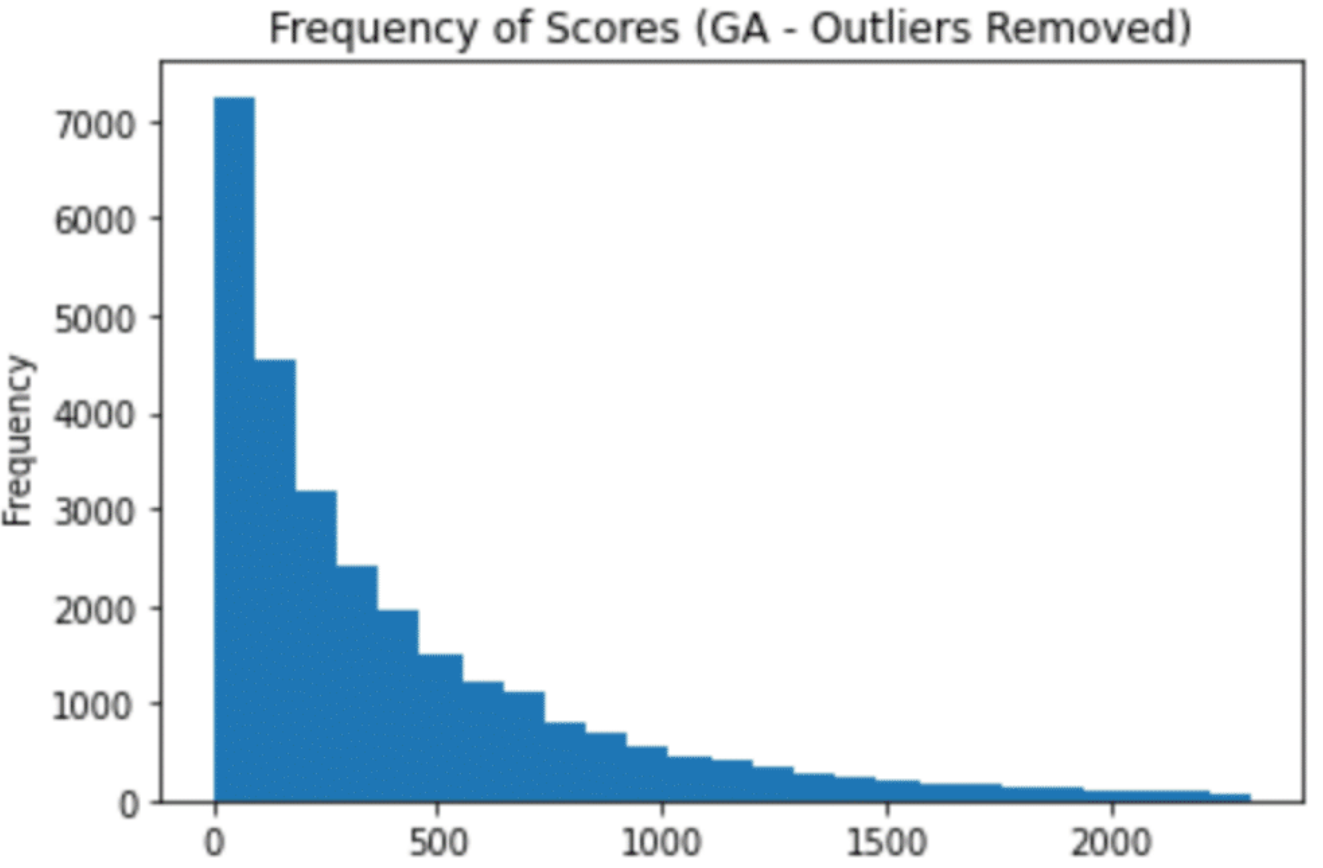}
    \caption{The frequency of scores in Georgia dataset.}
    \label{GAstat}
\end{figure}

\begin{figure}[t]
    \centering
    \includegraphics[scale=0.35]{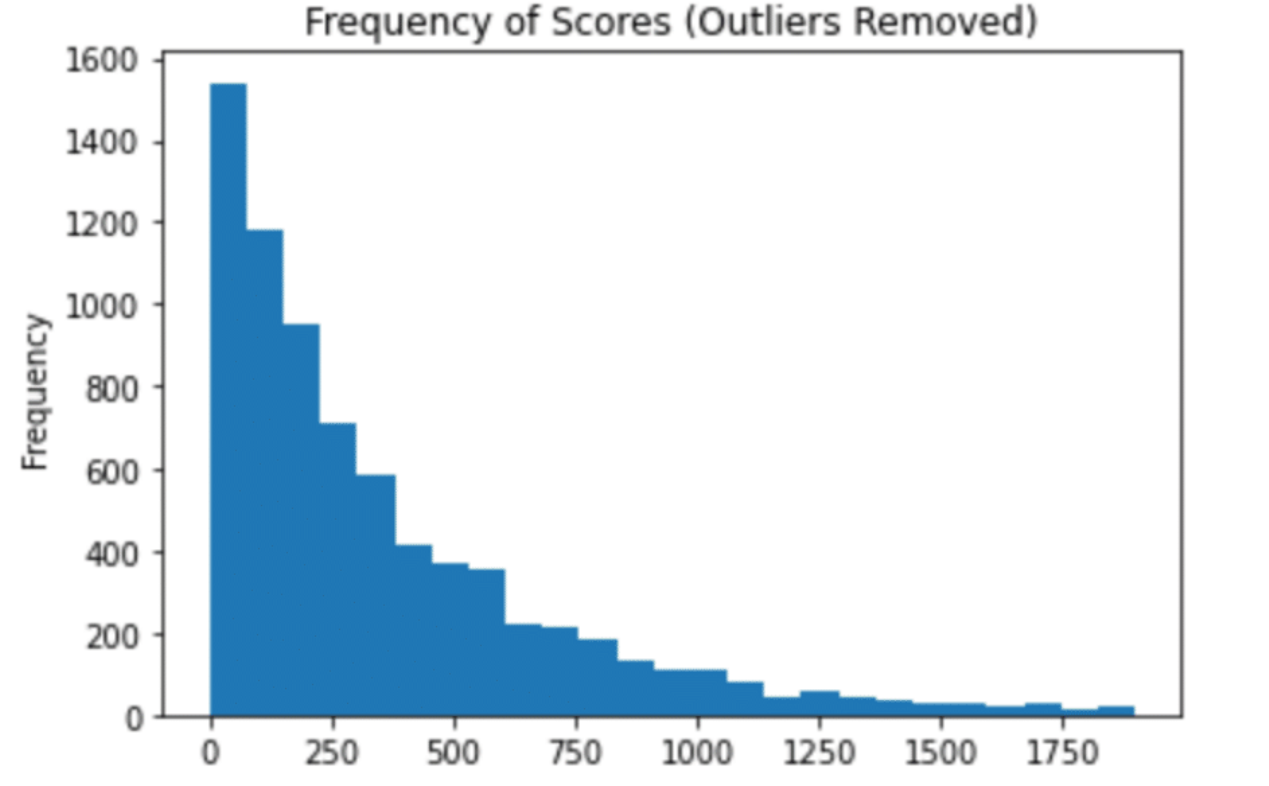}
    \caption{The frequency of scores in New Mexico dataset.}
    \label{NMstat}
\end{figure}

The following provides a brief description of some of these variables:
\begin{itemize}
\item Location: Denotes the location of the center of a $600m^2$ grid block where the users measure the connection speed.

\item Download speed: Represents the measured rate of data transfer from the server to the user's device. This is a critical metric as it affects the speed of web page loading and file downloading, and hence is quite noticeable to the user. Download speeds are reported as an average in the grid location. 

\item Upload speed: Corresponds to the measured data transfer rate from the user's device to the server. This metric is crucial for tasks like video calls, cloud file uploads, and user-generated live streaming. Upload speeds are reported as an average in the grid location. 

\item Number of tests at each location: This refers to the number of times tests were conducted in a given grid block. More tests tend to indicate a more reliable measurement score. However, in many locations, only a single test is conducted.

\item Number of devices used for testing: This refers to the variety of devices used for testing in a certain grid block, as the performance can vary across different mobile phones and mobile providers.

\item Score: This variable combines both the download and upload speeds to provide a holistic view of the connection
bandwidth performance.
\end{itemize}
We have mobile connection data from the states of Georgia in the United States Southeast and New Mexico in the United States Southwest. The geographic scatter plot of the connection scores for each dataset is shown in Figure~\ref{s1}. There are 28,587 data points in the Georgia dataset. The mean score across grid blocks is 460.067, the min score is 0.056, and the max score is 12421.537. The standard deviation of the score is 616.3. The frequency of scores is shown in Figure~\ref{GAstat}. There are 7,579 data points in the New Mexico dataset. The mean score across grid blocks is 388.128, the min score is 0.174, and the max score is 13202.754. The standard deviation of the score is 504.5. The frequency of scores is shown in Figure~\ref{NMstat}.

\begin{figure}[t]
    \centering
    \includegraphics[scale=0.315]{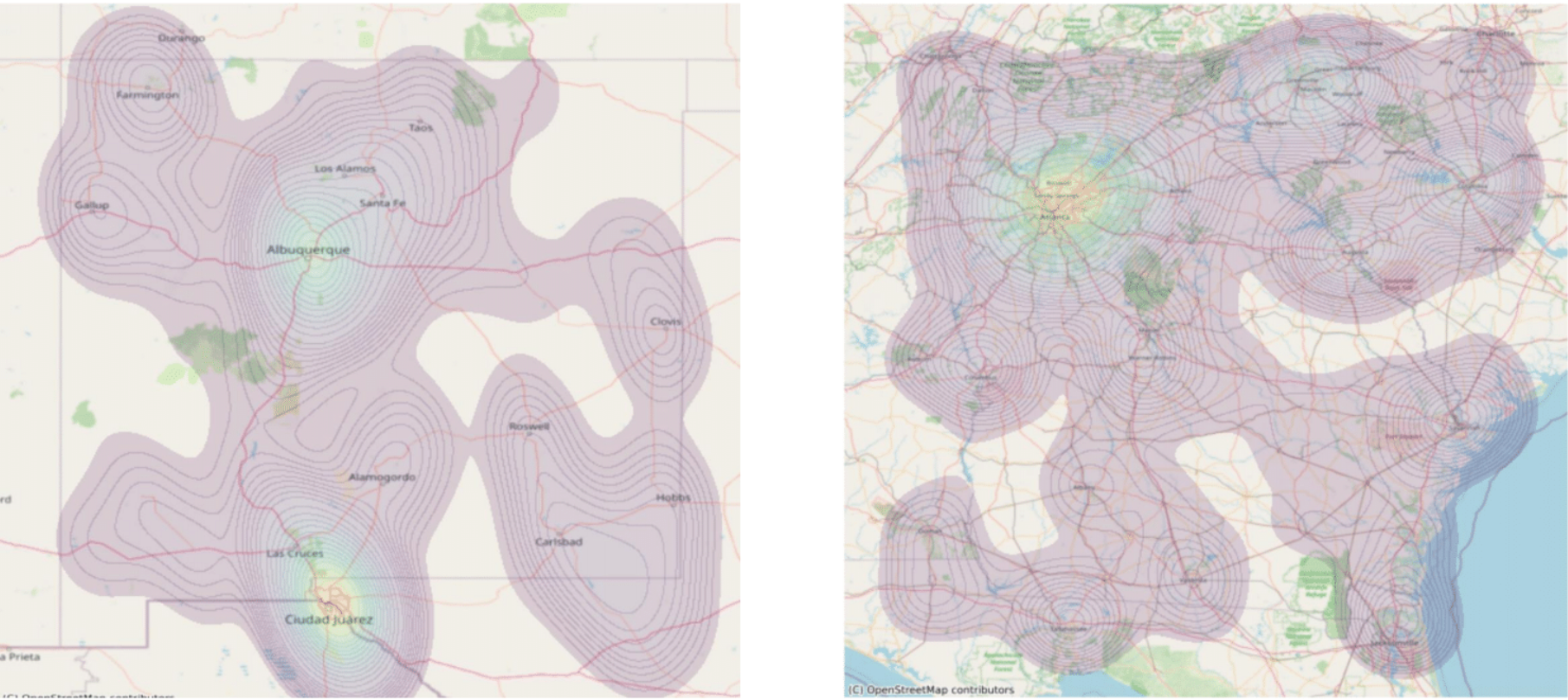}
    \caption{Kernel density estimation of Georgia and New Mexico dataset.}
    \label{s2}
\end{figure}

\section{Methodology}
We first employ standard kernel density estimation \cite{rosenblatt1956remarks} to generate a heatmap depicting the mobile connection quality across two distinct geographical areas, namely Georgia and New Mexico. These heatmaps, as displayed in Figure \ref{s2}, offer an illustrative representation of connection quality, enabling us to visually assess the spatial distribution of mobile connectivity in these regions.

However, upon inspection of the resulting heatmaps, it becomes evident that the standard kernel density estimation approach does not provide satisfactory coverage in certain rural areas. There are locations where the recorded measurements are notably sparse, resulting in large regions with no substantial estimations. Given that our study has a particular interest in understanding and improving the mobile connection quality in these rural and underserved areas, this observation underscores a significant limitation of the simple kernel density estimation method.

The apparent lack of estimations in the rural regions does not necessarily indicate an absence of mobile connectivity; rather, it highlights the deficiency of data in these areas, potentially due to their sparse population or challenging geographical conditions. It is precisely these areas where there is a need for robust and accurate analytical methods. Given that conventional kernel density estimation falls short in this regard, our observations demonstrate the necessity for adopting more sophisticated, data-driven techniques.

The more advanced methodologies should ideally be able to compensate for the sparse data in rural areas and provide reliable estimations, thereby offering a more comprehensive view of the mobile connectivity landscape. Moreover, this situation also points towards the need for further data collection efforts in such under-represented regions, which can help in refining our analysis and augmenting the efficacy of our predictive models.

\subsection{Gaussian Process Regression (Kriging)}
A Gaussian process (GP) is a collection of random variables, any finite number of which have a joint Gaussian distribution. This powerful Bayesian methodology can be used for various tasks such as regression, classification, and optimization. Gaussian processes are completely determined by their mean and covariance functions, and they provide a probabilistic, non-parametric approach to modeling data, meaning they can flexibly model complex relationships without relying on a fixed functional form.

Denote the output of the Gaussian process by $f(\mathbf{x})$, where $\mathbf{x}$ is the input. We can then express a Gaussian process as:
\begin{equation}
f(\mathbf{x}) \sim \mathcal{GP}(m(\mathbf{x}), k(\mathbf{x}, \mathbf{x'})),
\label{eq:GP}
\end{equation}
where $m(\mathbf{x})$ is the mean function, and $k(\mathbf{x}, \mathbf{x'})$ is the covariance function. They determine the properties of the functions drawn from the GP.

Given a set of $n$ observations $\mathbf{y} = [y_1, \dots, y_n]^T$ at locations $\mathbf{X} = [\mathbf{x}_1, \dots, \mathbf{x}_n]^T$, we can express the joint distribution of these observations as follows:
\begin{equation}
\mathbf{y} \mid \mathbf{X} \sim \mathcal{N}(m(\mathbf{X}), K(\mathbf{X}, \mathbf{X})).
\end{equation}
Here, $m(\mathbf{X})$ is a vector where the $i$-th entry is $m(\mathbf{x}_i)$, and $K(\mathbf{X}, \mathbf{X})$ is a covariance matrix where the entry at the $i$-th row and $j$-th column is $k(\mathbf{x}_i, \mathbf{x}_j)$.

Given the observed data, we can make predictions for new inputs $\mathbf{x}_*$ using the posterior predictive distribution:
\begin{align}
f\left(\mathbf{x}_* \mid \mathbf{y}, \mathbf{X}, \mathbf{x}_*\right) \sim \mathcal{N}\left(\bar{f}_*, \operatorname{cov}\left(f_*\right)\right),
\end{align}
where $\bar{f}_*$ is the predictive mean and $cov(f_*)$ is the predictive covariance.

The Gaussian process offers several advantages, such as providing a measure of uncertainty (via the predictive variance) in addition to point estimates and the flexibility of specifying different covariance functions. However, they also have some drawbacks. Its computational complexity scales cubically with the number of observations, which makes it difficult to tackle large datasets.

In our problem, we denote the inputs and response by $x\in\mathbb{R}^2$ and $y\in\mathbb{R}$, respectively. we utilize the kernel
\begin{align}
 k(\mathbf{x}, \mathbf{x'}) = \sigma^2 \exp(-(\mathbf{x}-\mathbf{x'})^T\text{diag}(\theta_1,\theta_2)(\mathbf{x}-\mathbf{x'})),
\end{align}
where $\theta_1, \theta_2$ and $\sigma$ are parameters to be trained through maximum likelihood. Once the model parameters are trained, we can use them for prediction.

\subsection{Kernel Regression}
Kernel regression is a non-parametric technique to estimate the conditional expectation of a random variable. The objective is to find a non-linear relationship between the input variable and the corresponding output. Kernel regression employs kernel functions, which allow it to capture more complex patterns than linear regression.

For a given dataset with inputs $\mathbf{x}$ and corresponding outputs $\mathbf{y}$, the kernel regression estimate $\hat{y}$ at a new input point $x$ is given by:
\begin{align} 
\hat{y}(x) = \frac{\sum_{i=1}^{n} K_h(\|x - x_i\|)y_i}{\sum_{i=1}^{n} K_h(\|x - x_i\|)}.
\label{KR}
\end{align}

Here, $K_h(u) = \frac{1}{h}K(\frac{u}{h})$ is a kernel function, and $h$ is a bandwidth parameter. The kernel function $K(\cdot)$ is often chosen to be a Gaussian kernel, though other choices are also possible. The bandwidth parameter $h$ controls the width of the kernel, and hence the smoothness of the estimated function.

In the Gaussian case, the kernel function $K(\cdot)$ is defined as:
\begin{align}
K(u) = \frac{1}{\sqrt{2\pi}}\exp\left(-\frac{1}{2}u^2\right).
\end{align}

Kernel regression estimates the conditional mean function without imposing a parametric form for the functional relationship between predictors and the outcome variable. It allows for flexible, data-driven model specification.

The bandwidth $h$ is a crucial parameter in kernel regression. If $h$ is too small, the estimate will be very rough, capturing too much noise in the data (overfitting). If $h$ is too large, the estimate will be too smooth, not capturing important patterns in the data (underfitting). The choice of an appropriate $h$ often involves cross-validation or some other form of bandwidth selection strategy.

\subsection{Self-tuning Bandwidth Kernel Regression}
Kernel regression, by its standard definition, utilizes a constant bandwidth $h$ that is applied uniformly across all data points. However, this one-size-fits-all approach might not be the most suitable in scenarios where the data exhibits spatial imbalances, as is the case in our study.

In regions where data points are sparse, the lack of neighboring points could potentially lead to unrepresentative averages and, consequently, inaccurate predictions. To overcome this, we propose the use of a larger bandwidth in such areas, thereby encompassing more points for computation and increasing the representativeness of the estimates.

To this end, we propose the self-tuning bandwidth in the kernel regression framework (STBKR). The self-tuning bandwidth mechanism adapts the bandwidth for each point based on its surrounding density of points. The formula for our self-tuning bandwidth kernel regression can be expressed as follows:
\begin{align}
\hat{y}(x) = \frac{\sum_{i=1}^{n} K_{h(x)}(\|x - x_i\|)y_i}{\sum_{i=1}^{n} K_{h(x)}(\|x - x_i\|)}.
\end{align}

In this formula, we let $h(x)=cR_k(x)^2$. Here, $c$ is a parameter that is determined by cross-validation, and $R_k(x)$ denotes the Euclidean distance from a given data point $x$ to its $k$-th nearest neighbor. Notably, in areas where data is sparsely distributed, the distance $R_k(x)$ will be larger, thereby leading to an increased bandwidth $h(x)$.

This strategy of self-tuning bandwidth effectively addresses the issue of spatial imbalance in the dataset. By allowing the bandwidth to adapt based on the local data density, it ensures a more representative sampling of neighbors, leading to more accurate and reliable predictions. Furthermore, the choice of $c$ through cross-validation aids in avoiding overfitting or underfitting, further strengthening the robustness of our regression model.

\begin{figure}[t]
\centering
\includegraphics[scale=0.35]{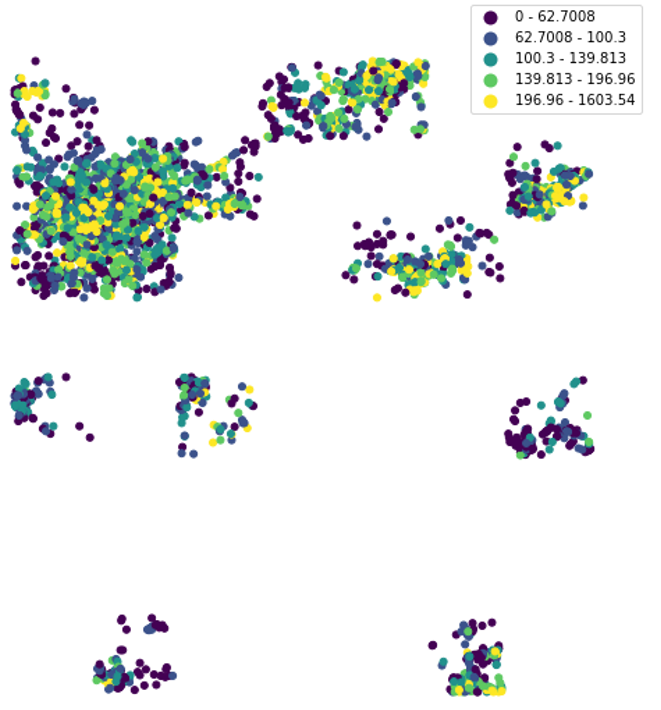}
\includegraphics[scale=0.35]{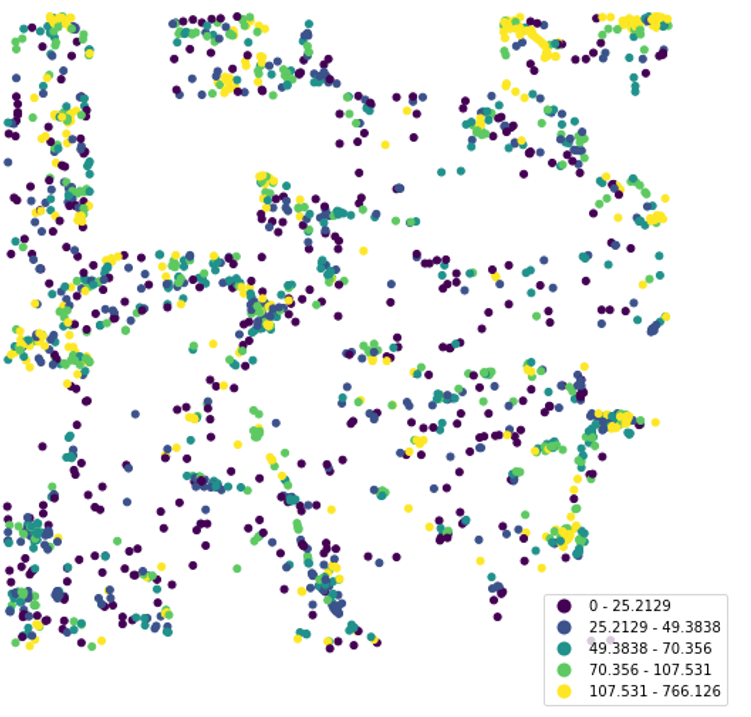}
\caption{Visualization of dense and sparse regions in the Georgia dataset.}
\label{heatmap}
\end{figure}

\begin{figure}[t]
\centering
\includegraphics[scale=0.5]{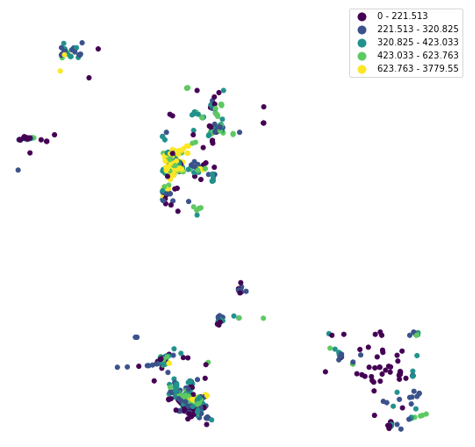}
\includegraphics[scale=0.5]{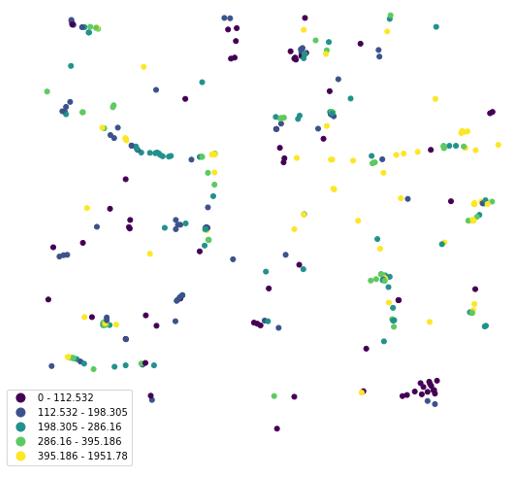}
\caption{Visualization of dense and sparse regions in the New Mexico dataset.}
\label{heatmap2}
\end{figure}

\subsection{Data Preprocessing and Efficient Computation}
The dataset under consideration presents a challenge due to its high variation and the fact that only a single measurement is taken at approximately half of the locations. This inherent variability in the data leads to the need for a variance reduction technique, specifically, the application of a $k$-nearest neighbor averaging method.

The $k$-nearest neighbor approach enables us to approximate the value at a given point $x$ by utilizing the mean of its $k$ nearest neighbors. Denote the $k$ nearest neighbors to a given point $x$ as $x_{n(x)_1},x_{n(x)_2},\cdots,x_{n(x)_k}$. The initial value $y$ at point $x$ is substituted by the average of these neighboring points as follows:
\begin{align}
y_{\text{avg}}(x) = \frac{1}{k}(y_{n(x)_1}+y_{n(x)_2}+\cdots+y_{n(x)_k}).
\end{align}
This averaging process inherently smoothens the dataset, helping to alleviate the issue of high variance.

However, a secondary challenge presents itself due to the size of the dataset. The computation of kernel regression predictions across the entire test set can take up to several hours, a prohibitively long duration for many applications. To mitigate this computational cost while retaining the prediction accuracy of our model, we resort to another variant of the k-nearest neighbor method, this time employing a weighted sum instead of summing over the entire training set as depicted in Equation \eqref{KR}.

With the incorporation of a kernel function $K_h(\cdot)$ that introduces a weighting based on the distance between the points, the predictive function $\hat{y}(x)$ is calculated as follows:
\begin{align}
\hat{y}(x) = \frac{\sum_{i=1}^{k} K_h(\|x - x_{n(x)_i}\|)y_{n(x)_i}}{\sum_{i=1}^{k} K_h(\|x - x_{n(x)_i}\|)}.
\label{WKR}
\end{align}

By adopting this approach, we can preserve the accuracy of the algorithm while dramatically reducing the computational time. This modification is instrumental in making the kernel regression computationally feasible for large datasets, thus striking a desirable balance between computational efficiency and predictive accuracy.

\section{Numerical Experiments}
\subsection{Experiment Setting}
We initiate our comparative study by partitioning the data into training and test sets with a split of $80\%$ and $20\%$ respectively. In our comparison, we feature our proposed self-tuning bandwidth kernel regression method (STBKR) and compare it with two baseline methods: The Gaussian process regression (GP) and fixed bandwidth kernel regression (FBKR). However, Gaussian process regression presents a computational challenge when dealing with large training data due to its requirement to compute the inverse of a massive matrix. We address this problem by sparsifying the training data through downsampling, followed by applying the Gaussian process to the thinned-out data.

To optimize the parameters $h$, $k$, and $c$ for the other two kernel regression methods, we implement a 5-fold cross-validation on the training set. The loss functions we apply in this study are Mean Absolute Error (MAE), Mean Squared Error (MSE), and Maximum Norm Error (MNE).

Furthermore, we segment our dataset into dense and sparse regions. To achieve this, we first divide the map into a $15\times 15$ uniform grid. Subsequently, we count the data points within each region. Regions with a count higher than the average are classified as dense, while the remaining ones are classified as sparse.

\begin{table}[t]
\caption{Errors in dense areas of Georgia dataset.}
\label{DG}
\vskip 0.15in
\begin{center}
\begin{small}
\begin{sc}
\begin{tabular}{lcccr}
\toprule
   & GP & FBKR & STBKR  \\
\midrule
MAE   & 245.62 & 83.03 & 81.6\\
MSE & 144105.22  & 20746.26 & 19594.86\\
MNE  &  5524.22  & 2334.2  &  2248 \\
\bottomrule
\end{tabular}
\end{sc}
\end{small}
\end{center}
\end{table}

\begin{table}[t]
\caption{Errors in sparse areas of Georgia dataset.}
\label{SG}
\vskip 0.15in
\begin{center}
\begin{small}
\begin{sc}
\begin{tabular}{lcccr}
\toprule
   & GP & FBKR & STBKR  \\
\midrule
MAE   & 178.69 & 102.1 & 102.9\\
MSE & 73267.2  & 47111.7 & 37534.83\\
MNE  &  2312.76  & 2399.2  &  1942.5 \\
\bottomrule
\end{tabular}
\end{sc}
\end{small}
\end{center}
\end{table}

Regarding the kernel regression methods, we tune the bandwidth magnitude $c$ and the number of neighbors $k$ separately in the dense and sparse regions. The motivation behind this approach is the reduced quantity of points in the sparse regions, which necessitates a larger bandwidth to include a comparable number of neighbors. This tailored treatment enhances the adaptability of our methods to the spatial distribution of the data, promoting better overall performance. 

Besides predicting the connection speed, we also tried a simpler task which is to predict whether an area's connection is served, underserved, or unserved. This 
distinction is useful in policy-making, as it has been
used to prioritize the order of broadband network expansion.

\begin{itemize}
    \item Served: Download speed $\ge 100$Mbps and upload speed $\ge 20$Mbps.
    \item Underserved: Download speed $\ge 25$Mbps and upload speed $\ge 3$Mbps and is not in served category.
    \item Unserved: The rest cases.
\end{itemize}

\subsection{Prediction}
In this experiment, we want to predict the score of each location in the test set accurately. First, we use 5-fold cross-validation on the training set to choose the best parameter $c$ and $k$ in dense regions and sparse regions for both datasets. The results are shown in Table \ref{tune1} and \ref{tune2}.

\begin{table}[t]
\caption{Parameters for dense and sparse areas in Georgia dataset.}
\label{tune1}
\vskip 0.15in
\begin{center}
\begin{small}
\begin{sc}
\begin{tabular}{lccr}
\toprule
   & FBKR & STBKR  \\
\midrule
Dense   & $k=5, c=0.01$ & $k=10, c=0.01$ \\
Sparse & $k=5, c=0.05$  & $k=5, c=0.05$\\
\bottomrule
\end{tabular}
\end{sc}
\end{small}
\end{center}
\end{table}
\begin{table}[t]
\caption{Parameters for dense and sparse areas in New Mexico dataset.}
\label{tune2}
\vskip 0.15in
\begin{center}
\begin{small}
\begin{sc}
\begin{tabular}{lccr}
\toprule
   & FBKR & STBKR  \\
\midrule
Dense   & $k=5, c=0.005$ & $k=5, c=0.05$ \\
Sparse & $k=5, c=0.075$  & $k=5, c=0.02$\\
\bottomrule
\end{tabular}
\end{sc}
\end{small}
\end{center}
\vskip -0.1in
\end{table}

For the Gaussian process, we do downsampling on training sets to reduce computational costs. To utilize the training data more efficiently and reduce the uncertainty in unmeasured areas, we split the map into a $15\times 15$ uniform grid and select training data uniformly in each grid. This makes the training data spread more uniformly and thus greatly reduces the uncertainty around the training set.

Then we compare the three methods separately in dense areas and sparse areas of the Georgia and the New Mexico dataset.

\begin{table}[t]
\caption{Errors in dense areas of New Mexico dataset.}
\label{DG}
\vskip 0.15in
\begin{center}
\begin{small}
\begin{sc}
\begin{tabular}{lcccr}
\toprule
   & GP & FBKR & STBKR  \\
\midrule
MAE   & 172.31 & 53.13 & 54.55\\
MSE & 80417.66  & 9224.37 & 8950.31\\
MNE  &  3038.29  & 1055.09  &  1049.89 \\
\bottomrule
\end{tabular}
\end{sc}
\end{small}
\end{center}
\end{table}

\begin{table}[t]
\caption{Errors in sparse areas of New Mexico dataset.}
\label{SG}
\vskip 0.15in
\begin{center}
\begin{small}
\begin{sc}
\begin{tabular}{lcccr}
\toprule
   & GP & FBKR & STBKR  \\
\midrule
MAE   & 115.3 & 77.34 & 77.55\\
MSE & 33754.41  & 31870.79 & 24624.49\\
MNE  &  1905.78  & 1172.39  &  1057.99 \\
\bottomrule
\end{tabular}
\end{sc}
\end{small}
\end{center}
\end{table}

Upon analysis of the experimental results, it becomes apparent that the Gaussian Process (GP) regression underperforms in regions of higher data density compared to those characterized by sparsity. This deficiency may be attributed to the employed downsampling technique, which uses a uniform distribution of training data across the spatial domain, thus failing to sufficiently represent areas of high data concentration. Furthermore, GP demonstrates much worse performances in comparison to kernel regression methods across both region types. This can be ascribed to GP's inherent limitation in handling large volumes of training data. Despite the allocation of $80\%$ of the total data to the training set, the downsampling process results in a mere $10\%$ of the data being used for training GP. In stark contrast, kernel regression methods exhibit the capacity to effectively utilize the entirety of the training set, leading to superior performance.

In a comparison of the standard kernel regression method and the proposed self-tuning kernel regression, our method consistently manifests superior performance in both regions across the datasets in question. While the Mean Absolute Error (MAE) remains relatively consistent between the two kernel regression methods, our self-tuning method exhibits a marked decrease in Mean Squared Error (MSE) and Maximum Norm Error ($l_{\infty}$ error). This indicates that our method offers enhanced predictive accuracy in those locations deemed more challenging, aligning with the primary objectives of our research.

\begin{figure}[t]
    \centering
    \includegraphics[scale=0.2]{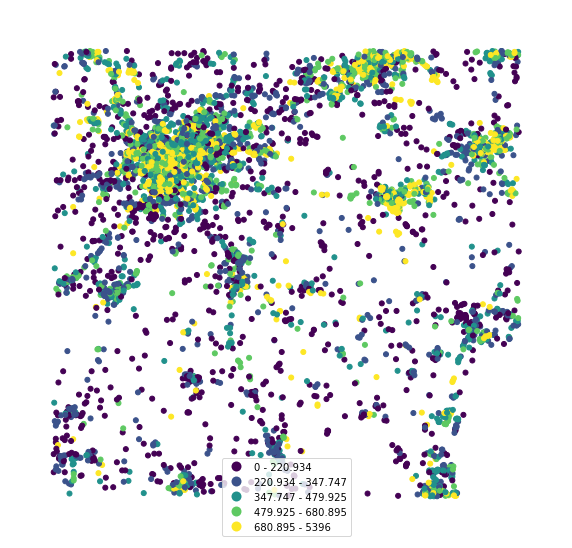}
    \includegraphics[scale=0.2]{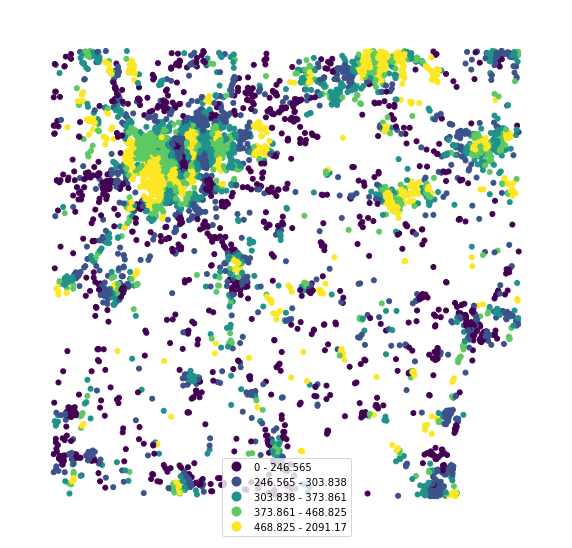}
    \includegraphics[scale=0.2]{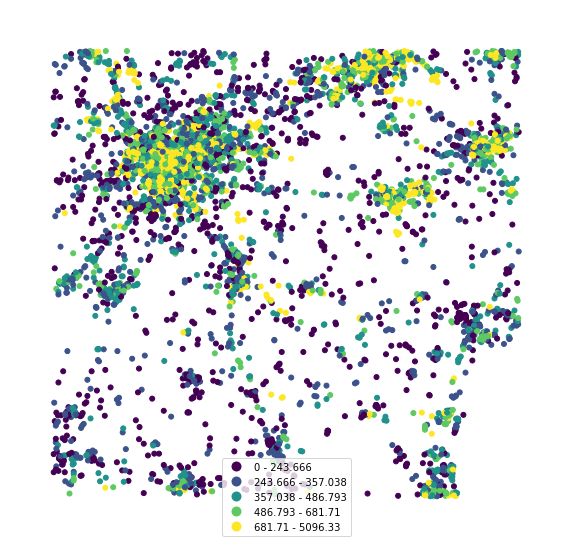}
    \includegraphics[scale=0.2]{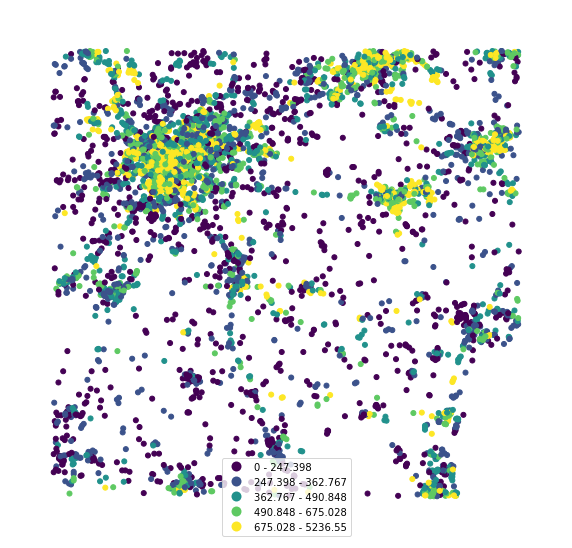}
    \caption{Top-left is the original test data after data averaging. Top-right is the prediction of GP. Bottom-left is the prediction of kernel regression. Bottom-right is the prediction of self-tuning kernel regression.}
    \label{heatmap}
\end{figure}

\begin{figure}[t]
    \centering
    \includegraphics[scale=0.2]{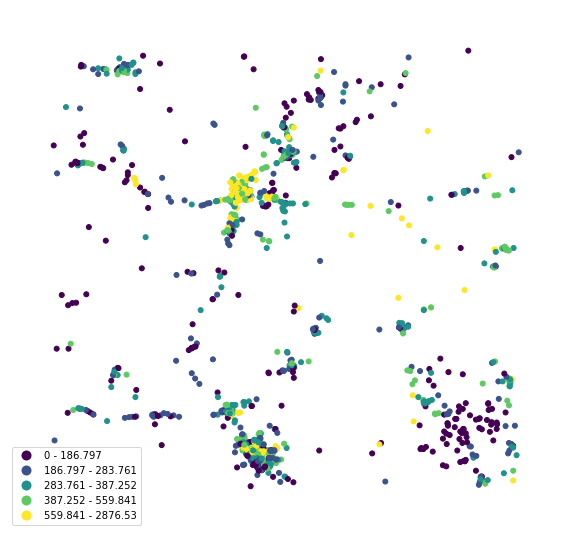}
    \includegraphics[scale=0.2]{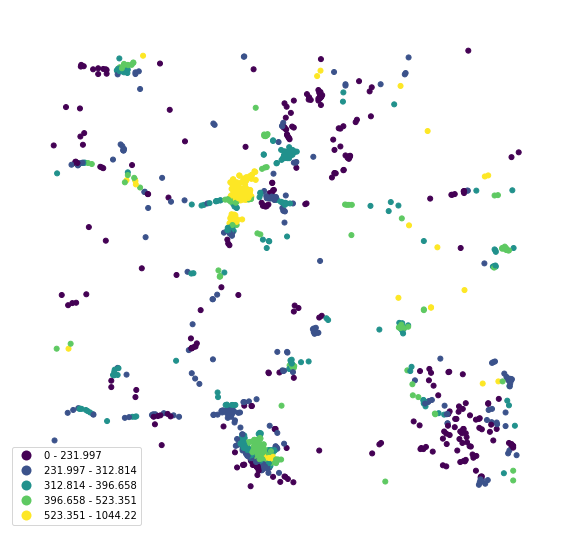}
    \includegraphics[scale=0.2]{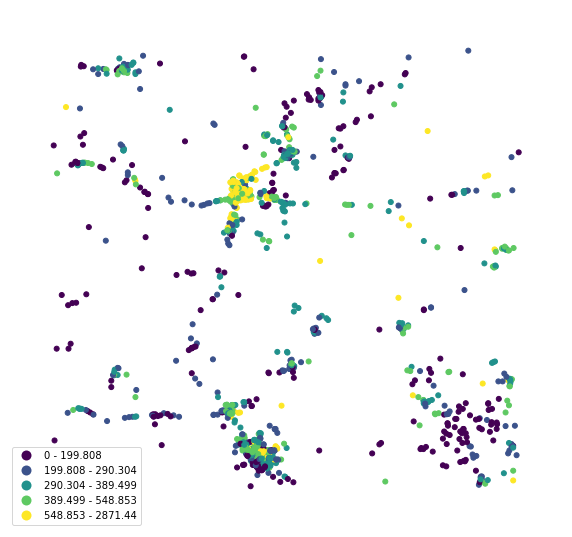}
    \includegraphics[scale=0.2]{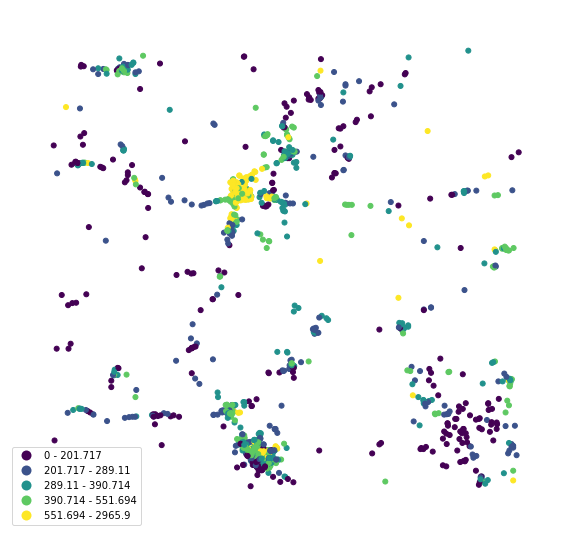}
    \caption{Top-left is the original test data after data averaging. Top-right is the prediction of GP. Bottom-left is the prediction of kernel regression. Bottom-right is the prediction of self-tuning kernel regression.}
    \label{heatmap}
\end{figure}

\subsection{Classification}
In this section, we classify locations into three types: served, underserved, and unserved regions. For the Georgia dataset, we have $48\%$ served, $49\%$ underserved, and $3\%$ unserved areas. In the New Mexico dataset, $21\%$ are served, $60\%$ are underserved, and $19\%$ are unserved. This shows significant differences in mobile connectivity between these two states.

We use the three methods discussed previously to predict both download and upload speeds. Each point in the test set is then classified based on these predictions. The final accuracy of this classification is provided below.

From the table \ref{CG} and \ref{CN}, we can see that our proposed method still generates a more accurate classification in all cases except the sparse region in the Georgia dataset. However, the difference is very small in that case, which does not obscure the overall advantage of our method.
\begin{table}[t]
\caption{Classification accuracy on Georgia dataset.}
\label{CG}
\vskip 0.15in
\begin{center}
\begin{small}
\begin{sc}
\begin{tabular}{lcccr}
\toprule
   & GP & FBKR & STBKR  \\
\midrule
Dense   & $52\%$ & $86.6\%$ & $88.3\%$\\
Sparse & $67.4\%$  & $86.2\%$ & $85.9\%$\\
\bottomrule
\end{tabular}
\end{sc}
\end{small}
\end{center}
\end{table}

\begin{table}[t]
\caption{Classification accuracy on New Mexico dataset.}
\label{CN}
\vskip 0.15in
\begin{center}
\begin{small}
\begin{sc}
\begin{tabular}{lcccr}
\toprule
   & GP & FBKR & STBKR  \\
\midrule
Dense   & $27.4\%$ & $88.7\%$ & $91.3\%$\\
Sparse & $20.8\%$  & $90.9\%$ & $93.3\%$\\
\bottomrule
\end{tabular}
\end{sc}
\end{small}
\end{center}
\end{table}

\section{Conclusion}
In this work, we focus on the cellular network quality estimation based on Ookla open dataset. We proposed a novel self-tuning bandwidth kernel regression method (STBKR) for the specific problem on two different datasets. The difficulty of the problem comes from data imbalance and high variation. We compared our method with two established approaches: the Gaussian process (GP) and the basic kernel regression. Our findings indicate that GP struggles with the dataset, especially in dense regions, whereas both kernel regression methods perform impressively due to their ability to employ the entire training set effectively.

Our method consistently shows superior performance over the other two baseline methods, particularly at locations with challenging conditions. This is further validated in our classification experiment, which seeks to categorize various locations based on connectivity conditions. Despite substantial differences in mobile connectivity across the datasets, our proposed method demonstrates superior classification accuracy.

To summarize, our self-tuning bandwidth kernel regression method offers a promising alternative for dealing with large datasets, delivering high predictive accuracy and effective classification performance across diverse regions. Its superiority over the Gaussian process regression and fixed bandwidth kernel regression methods substantiates its potential as a valuable tool in data analysis.

\nocite{langley00}

\bibliography{example_paper}
\bibliographystyle{icml2023}



\end{document}